\documentclass[proof]{WileyASNA-v1}

\newcommand{\xmm}{\textit{XMM-Newton}}

\makeatletter
\def\blfootnote{\gdef\@thefnmark{}\@footnotetext}
\makeatother

\articletype{Original Paper}
\received{<day> <Month>, <year>}
\revised{<day> <Month>, <year>}
\accepted{<day> <Month>, <year>}

\begin{document}

\title{X-ray irradiation of three planets around Hyades star K2-136}

\author[1,2]{Jorge Fernandez Fernandez*}
\author[1,2]{Peter J. Wheatley}

\authormark{J. Fernandez \textsc{et al.}}

\address[1]{\orgdiv{Centre for Exoplanets and Habitability}, \orgname{University of Warwick}, \orgaddress{Gibbet Hill Road, Coventry CV4 7AL, \country{UK}}}
\address[2]{\orgdiv{Department of Physics}, \orgname{University of Warwick}, \orgaddress{Gibbet Hill Road, Coventry CV4 7AL, \country{UK}}}

\corres{*Jorge.Fernandez-Fernandez@warwick.ac.uk}

%\presentaddress{Department of Physics, University of Warwick, Gibbet Hill Road, Coventry CV4 7AL, UK}

% =========================================

\abstract{
We study the X-ray irradiation and likely photoevaporation of the three planets around the star K2-136. These are the Earth-sized K2-136 b, the mini-Neptune K2-136 c, and the super-Earth K2-136 d. \xmm{} observations of the star indicate an X-ray luminosity of $(1.18\pm0.1)\times10^{28}$ erg s$^{-1}$ in the range 0.15 to 2.4 keV, resulting in an activity of $L_X/L_{\rm bol}=(1.80\pm0.68)\times10^{-5}$. The evaporation past of the planets were modelled using the XUV stellar tracks by \citet{johnstone-2020}, the energy limited mass loss formulation \citep{lecavelier-des-etangs-2007, erkaev-2007}, and the thermal evolution formulation by \citet{lopez-fortney-2014}. Our results suggest that planets b and d are most probably purely rocky and have been stripped of their envelopes. On the other hand, planet c likely still has an envelope consisting of 0.8\% of the total mass, with its core being relatively large (2.3 $R_{\rm E}$).
}


\jnlcitation{\cname{%
\author{J. Fernandez Fernandez}, and
\author{P. J. Wheatley}} (\cyear{2021}), 
\ctitle{The photoevaporation of three planets around Hyades star K2-136}, \cjournal{Journal}, \cvol{Vol}.}

\maketitle

% =========================================

\section{Introduction}

The Kepler mission has uncovered a surprisingly large population of planets between Earth and Neptune in size. Their radii have been observed to follow a bimodal distribution \citep{fulton-2017}, with twin peaks at 1.3 and 2.4 $R_{\rm E}$, and a gap at 1.8 $R_{\rm E}$. Whilst the first group is populated by dense rocky planets, the second one can be explained by the additional presence of H/He gaseous envelopes that can double a planet's radius and yet comprise less than a percent of its mass. This phenomenon can be observed as a valley on the radius-period parameter space in Figure \ref{fig:planets-valley}. There is evidence that points to stellar high-energy radiation being the main driver for this behaviour \citep[e.g.][]{owen-wu-2013, owen-2019}, although other channels such as core-powered mass loss have also been suggested  \citep[e.g.][]{core-powered-mloss}.

The X-ray and extreme ultraviolet radiation (together XUV) that stars emit is absorbed by the upper layers of the atmospheres of close-in exoplanets. This heats the upper atmosphere driving a hydrodynamic wind that overflows the Roche lobe and escapes the planet. Gradual mass loss sculpts the observed planet populations, particularly during the first 100 Myr when the XUV emission is the strongest, stripping the envelopes from planets that are too close or too light. There is also evidence that EUV radiation exerts significant influence on envelopes on much-longer Gyr timescales \citep{king-2021}. 

Observations, however, only provide a snapshot of a planet's current evolutionary state. One can attempt to model its current physical properties but its history can be hard to constrain due to uncertainty in the stellar rotation and hence X-ray luminosity history. A straightforward solution to break the degeneracy is to study multiplanetary systems: where all the planets in the system share the same the XUV irradiation history.  Their current states can then be used to constrain each other's pasts, especially if the planets are found on both sides of the evaporation valley \citep{campos-estrada-20}.

\section{The target}

K2-136 is a K-type star located $59$ parsecs away (as determined by Gaia EDR3). It is a member of the Hyades cluster, making it around 600 Myr old \citep{perryman-1998}, although more recent estimates go as high as 800\,Myr \citep{brandt-huang-2015, david-hillenbrand-2015}. In this work, we adopt an age of $700\pm100$ Myr. The star also hosts a three-planet system: the Earth-sized K2-136 b, the mini-Neptune K2-136 c, and the super-Earth K2-136 d, which are all situated within 0.2 AU of their host star \citep{livingstone-2018, mann-2017, ciardi-2018}. \citet{ciardi-2018} also reports the existence of a candidate stellar-mass companion, a M7/8V dwarf, at a projected separation of 40 AU, although follow-up observations are still needed to confirm the stars are gravitationally bound to each other.

\subsection{Planetary system}

\begin{figure}[t]
\centerline{
\includegraphics[scale=0.5]{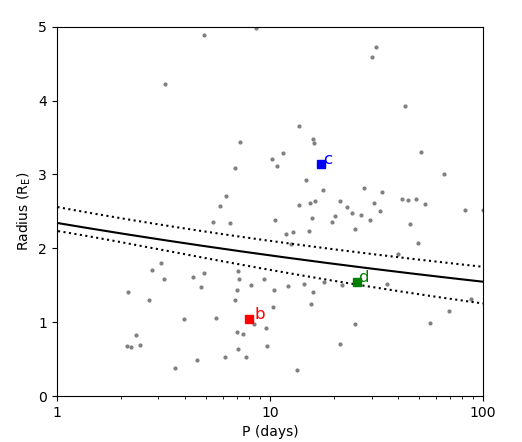}
}
\caption{Radius against orbital period plot of the exoplanet population from \citet{van-eylen-2018} alongside their fit for the evaporation valley (solid black line, uncertainties as two dotted lines). The three K2-136 planets are plotted in red (b), blue (c), and green (d). \label{fig:planets-valley}}
\end{figure}

The three discovery papers of the planetary system \citep{livingstone-2018, ciardi-2018, mann-2017} agree within $1\sigma$ in their estimates for the periods and radii of the three planets. These are $R_b=1.05\pm0.16$ R$_{\rm E}$ and $P_b=7.975$ days for planet b, $R_c=3.14\pm0.36$ R$_{\rm E}$ and $P_c=17.307$ days for planet c, and $R_d=1.55\pm0.24$ R$_{\rm E}$ and $P_d=25.575$ days for planet d. All periods have uncertainties of less than a percent. These are put in context with the evaporation valley on Figure \ref{fig:planets-valley}. It is interesting how the middle planet (c) is the one that remains above the valley whereas the one furthest away is not. This points to planet c having a much heavier core than d in order to avoid having its envelope stripped.

\begin{figure}[t]
\centerline{
\includegraphics[scale=0.55,trim={1.5cm 0 0 0}]{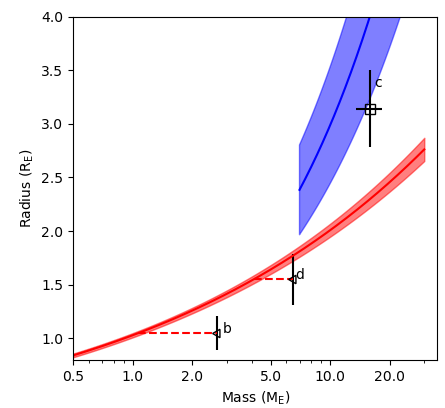}
}
\caption{\citet{otegi-2020} mass-radius relations for rocky (red) and volatile-rich (blue) planets, alongside the three K2-136 planets.\label{fig:planets-mr}}
\end{figure}

The planet masses have also been recently measured by \citet{mayo-2021}; with a direct estimate for the mini-Neptune ($15.9\pm2.4$ M$_{\rm E}$) and upper limits for the other two ($<2.67$ and $<6.47$ M$_{\rm E}$ for b and d, respectively). We make use of the mass-radius relationships by \citet{otegi-2020} to obtain mass estimates for b and d, and to put planet c in context (Figure \ref{fig:planets-mr}). 
The small radii of planets b and d place them below the evaporation valley, probably with purely solid cores with no envelope. Their masses are estimated to be $1.07^{+0.67}_{-0.46}$ M$_{\rm E}$ and $4.09^{+2.63}_{-1.80}$ M$_{\rm E}$ respectively. In contrast, the mass of planet c indicates a large core, and its location above the valley points to the presence of an envelope. As its envelope likely contributes less than a percent of its mass, we fit its mass to the rocky relation and obtain an estimate for the core size of $R_{c,core} = 2.30\pm0.13$ R$_{\rm E}$, hence its envelope would comprise 25\% of the total radius.

\subsection{The Hyades}
\begin{figure*}[t]
\centerline{
\includegraphics[scale=0.7,trim={0 0 0 0.55cm},clip]{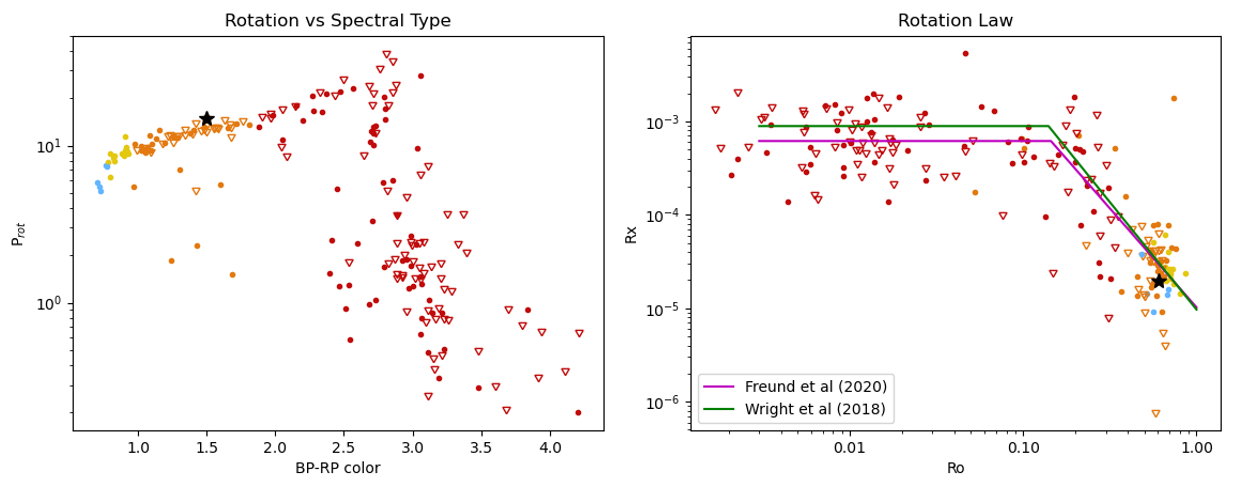}
}
\caption{Left: rotation period of Hyades members versus colour index (Gaia blue and red magnitudes, BP--RP). Right: X-ray luminosity against Rossby number for Hyades members. The rotation-activity relation by \citet{wright-2018} is plotted in green and the relation for Hyades members  by \citet{freund-2020} in magenta. Data have been taken from \citet{freund-2020}. K2-136 is plotted as a black star in both panels. \label{fig:hyades}}
\end{figure*}

The Hyades is a well studied nearby open cluster, with several extensive X-ray studies \citep{stern-1981, micela-1988, collura-1993} as well as rotational surveys \citep{delorme-2011, douglas-2019}. \citet{freund-2020} compiled 1100 Hyades members from several studies, both from the core of the cluster and its recently discovered tidal tails. About 130 of these cluster members have measurements for both the rotation periods and the X-ray luminosity. The data presents a tight rotation-spectral type relation for F, G, and K-dwarfs (Figure \ref{fig:hyades}, left panel), with M-dwarfs being much more scattered. According to \citet{johnstone-2020}, M-dwarfs retain their initial distribution of rotation rates for longer before settling into a single mass-dependent value, joining the tight relation of F,G, and K-type stars. K2-136 fits well on the tight relation. The stray K-type stars at faster rotation rates below the tight relation are likely close binaries (and indeed we see they are offset from the main sequence in an H-R diagram). 

\section{Observations}
\label{sec:obs}
We observed the target with the \xmm{} telescope in November 2018 for 43 kiloseconds. Data reduction was performed on the EPIC-pn data, which has an operative energy range of 0.15 to 12 keV. The spectrum was modelled with a TBABS model \citep{tbabs} to account for interstellar absorption as well as an APEC model \citep{apec} for the emission spectrum from a collisionally-ionized diffuse plasma. Solar abundances by \citet{abund-aspl} were used. The hydrogen column density was set to $n_H = 10^{18}$ cm$^{-2}$, as estimated by \citet{redfield-linksy-2001} for the Hyades. Furthermore, the ROSAT-range flux was estimated by extrapolating the model down to 0.1 keV, and the EUV flux from the relations by \citet{king-2018}.

\section{Data analysis}
\label{sec:fluxes}
\begin{figure}[t]
\centerline{
\includegraphics[scale=0.55,trim={0 0 0 0},clip]{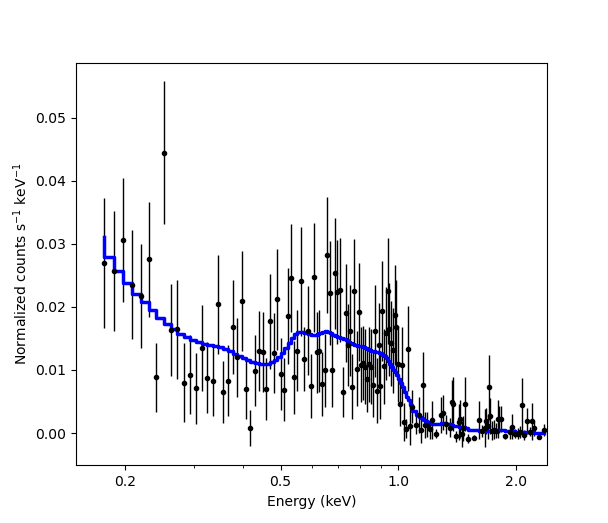}
}
\caption{X-ray spectrum of K2-136 as observed with the \xmm{} telescope. The model fitted (blue) is described in Sections \ref{sec:obs} and \ref{sec:fluxes}.\label{fig:spectrum}}
\end{figure}

Since the source is only detected up to 2.4 keV, only the range 0.15 to 2.4 keV was used for model fitting. Due to the low number of counts per bin, we use the C statistic as the fit parameter \citep{cash-1979}. We find that a three-temperature APEC model works best for our data, with temperatures $0.14^{+0.05}_{-0.06}$, $0.29^{+0.17}_{-0.06}$, and $0.97^{+0.17}_{-0.14}$ keV. The spectrum is plotted in Figure \ref{fig:spectrum}.

Integrating the model yields a flux of $F_X = 2.83\pm0.25 \times 10^{-14} $ erg cm$^{-2}$ s$^{-1}$, resulting in a luminosity of $1.18 \pm 0.1 \times 10^{28}$ erg s$^{-1}$ in the range 0.15 to 2.4 keV. Extrapolating down to an energy of 0.1 keV yields a ROSAT-range luminosity of $1.32\pm0.12\times10^{28}$ erg s$^{-1}$.

Some of this X-ray flux is likely to arise from the close-by M-dwarf companion. We estimate this from the measured 2MASS fluxes of the candidate companion, which \citet{ciardi-2018} find to be $J = 14.1\pm0.1$, $H = 13.47\pm0.04$, and $Ks= 13.03\pm0.03$. Interpolating with the stellar quantity sequences by \citet{boltable}, we estimate its bolometric luminosity to be $L_{\rm bol} = 2.00\pm0.35\times10^{30}$ erg s$^{-1}$, about $5\times10^{-4}$ times lower than that of K2-136. Since \citet{wright-2011} finds that stars tend to be saturated at an X-ray activity of $log_{10} R_X = -3.13\pm0.08$, and we assume that such a low mass star remains X-ray saturated even after 800 Myr \citep{johnstone-2020}, we can estimate its X-ray luminosity to be $L_X \approx 1.48\pm0.38\times10^{27}$ erg s$^{-1}$, about 10\% of the flux from K2-136 measured in this work, which is comparable to the uncertainty in our measurement.

Finally, we estimate the EUV flux using the relations by \citet{king-2018} and thus obtain a value of $L_{\rm EUV} = 2.38\pm0.17\times10^{28}$ erg s$^{-1}$ in the range 0.0136 to 0.15 keV. This results in a combined XUV luminosity of $L_{\rm XUV}=3.7\pm0.2\times10^{28}$ erg s$^{-1}$, three times stronger than X-rays alone.

Plotting the star on the rotation-activity relation with its cluster siblings (Figure \ref{fig:hyades}, right panel) indicates good agreement with the other Hyades K-dwarfs. It is also worth noting the large scatter in activity around the fit, up to one order of magnitude. It is yet to be explained whether this originates from random variability or intrinsic tracks for each star.

\section{Evolution modelling}
In order to explore the pasts of the planetary envelopes, three ingredients are necessary: the XUV emission track of the host star as it spins down, a formulation for the mass loss that updates the envelope mass on each time step, and a description of the planetary interior that relates the envelope mass fraction to its size, vital for recalculating the envelope radius after mass is lost on each time step.

\subsection{Stellar tracks}
The XUV emission of a star is tied to its Rossby number, the ratio between rotation period and convective turnover time \citep[e.g.][]{wright-2011}. Stars spin down with age, which results in a decrease of the star's emission with time, eventually diminishing its capability to evaporate atmospheres past a few Gyr.

\begin{figure*}[t]
\centerline{
\includegraphics[scale=0.55]{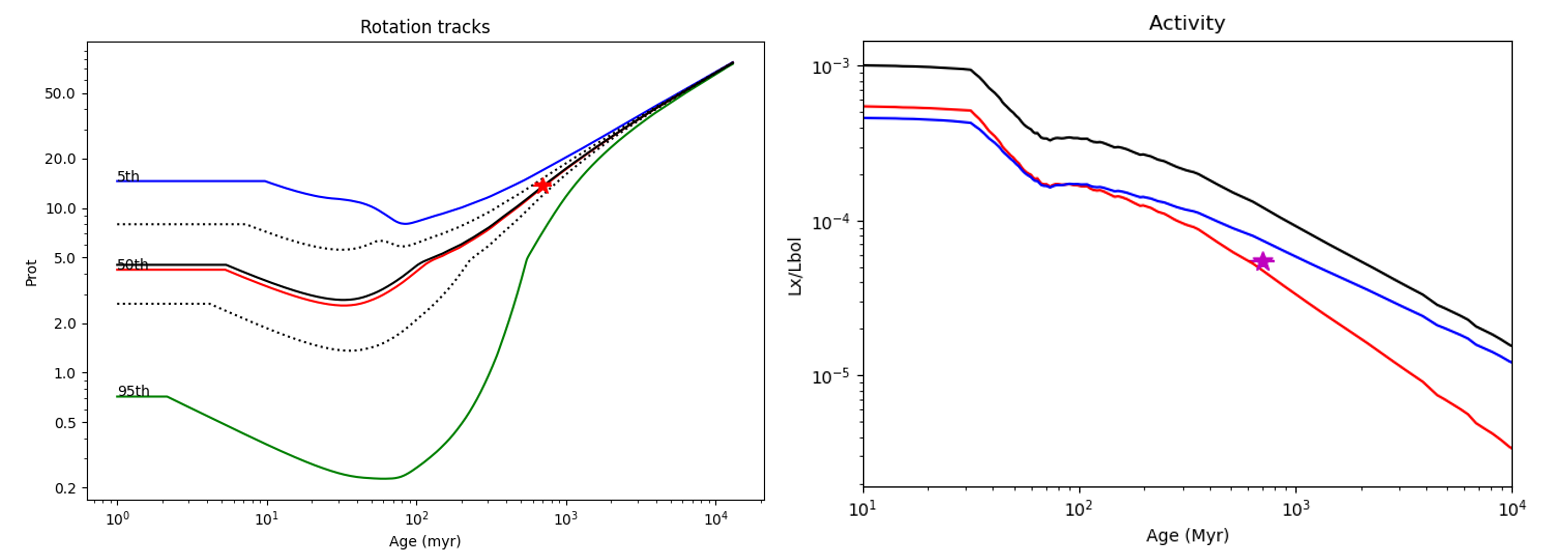}
}
\caption{Stellar tracks from the model by \citet{johnstone-2020}. Left: possible rotation tracks for a 0.7 $M_{\odot}$ star given a range of initial rotation rates; 5th percentile (blue), 50th percentile (red), and 95th percentile (green). The track that best fits K2-136 is the solid black line, with the uncertainties set by the two dotted black lines. Right: X-ray activity $R_X = L_X / L_{\rm bol}$ (black line). The X-ray (red) and EUV (blue) tracks are plotted separately as well. K2-136 is plotted as a magenta star.\label{fig:tracks}}
\end{figure*}

We make use of the physically motivated X-ray tracks proposed by \citet{johnstone-2020}, which model the radiative core and convective envelope of F, G, and K-type stars as two independently rotating solid components. Each component has certain angular momentum affected by torques of different origins (stellar wind, momentum exchange, and PMS spin-up), which can be used to predict the star's rotation and XUV dependence on age. These tracks are plotted in Figure \ref{fig:tracks} (left panel) for a star of the mass of K2-136 ($0.7$ M$_{\odot}$) and a distribution of initial rotation rates. K2-136 seems to fit well with the average rotation track (50th percentile).

\subsection{Thermal evolution}
We adopt the envelope structure model described by \citet{lopez-fortney-2012}, which models the cooling and contraction of adiabatic envelopes assuming a hot-start model. The planet's energy budget is tracked by taking into account heat from different sources, such as stellar radiation, radioactive decay from heavy elements, and thermal inertia from the isothermal rocky core. Although they begin their models at 1 Myr, the hot-start assumption might result in over-inflated radii (>10 R$_{\rm E}$) for low-mass planets at early ages. Despite that, these envelopes cool down and contract quickly and should be insensitive to the choice of initial entropy by the age of 10 Myr. Computed model tables and polynomial fits are provided by \citet{lopez-fortney-2014}, which only run from 100 Myr to 10 Gyr, however. Regardless, we choose a lower age limit of 10 Myr in our simulations as most of the mass loss occurs before 100 Myr.

\subsection{Mass loss}
The photoevaporation is modelled using the energy-limited approach \citep{watson-1981, lecavelier-des-etangs-2007, erkaev-2007}. This balances the incident energy flux from the star against the energy necessary to displace matter from the planet’s upper atmosphere to its Roche lobe. The mass loss rate is given by
\begin{equation}
    \dot{M} = \frac{\beta^2 \eta \pi F_{\rm XUV} R^3_p}{G K M_p},
    \label{eqn:massloss}
\end{equation}
where $M_p$ and $R_p$ are the mass and radius of the planet, respectively, $F_{\rm XUV}$ is the incident combined X-ray and EUV flux into the planet, $\beta$ is the ratio between the radii at which the atmosphere becomes optically thick for XUV wavelengths and for visible wavelengths ($R_{\text{\rm XUV}}/R_p$), $\eta$ is the energy efficiency of the process, and $K$ accounts for the potential energy difference between the surface and the height of the Roche-lobe (see \citet{erkaev-2007} for more details). This approach is attractively simple, although it does mask more complex physics behind the $\eta$ and $\beta$ parameters.

Generally, constant canonical values for these two parameters are used, but there more sophisticated formulations as well. For instance, \citet{salz-2016} perform hydrodynamic simulations to derive scaling laws for the efficiency and the XUV radius based on both the gravitational potential of a planet and its irradiation level. In this work, we adopt a canonical efficiency of $\eta = 15\%$ and a XUV radius ratio $\beta = 1$ \citep{king-2018}, and we compare with the formulation by \citet{salz-2016} for $\beta$.

\subsection{Simulation results}

Planets c and d were de-evolved from their current states back to 10 Myr. This assumes that planet d just finished losing its now-missing envelope, which sets an upper limit for the total mass lost. Using an efficiency of $\eta=15$\% and a XUV radius ratio of $\beta=1$, at 10 Myr we obtain envelope mass fraction of 0.83\% for planet c, expanding from 3.14 to $\sim$3.8 R$_{\rm E}$, and 0.12\% for planet d, inflating its radius from 1.55 to $\sim$2.15 R$_{\rm E}$. 

\begin{figure}[t]
\centerline{
\includegraphics[scale=0.25]{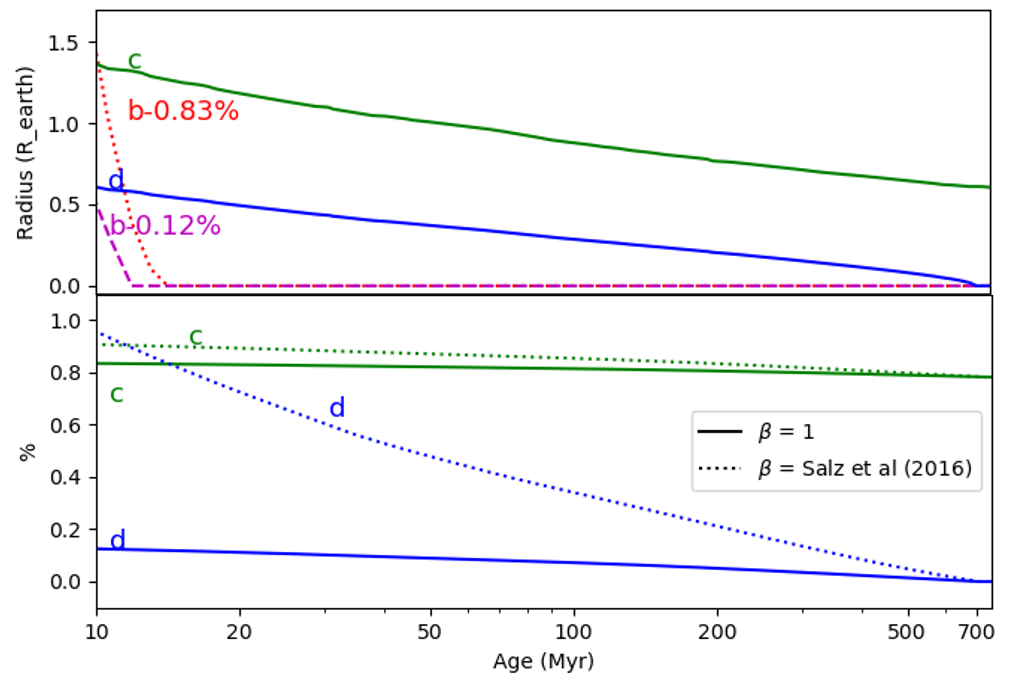}
}
\caption{Top: planet radius against age. Planets c (green line) and d  (blue) were de-evolved from their current age of 700 Myr. Using their envelope mass fractions at 10 Myr, planet b was evolved forward to 700 Myr, with initial fractions of $M_{env}/M_p=0.83\%$ (red dotted line) and $M_{env}/M_p=0.12\%$ (magenta dashed line). The envelope loss timescales for planet b were estimated to be $\sim$2 and $\sim$4 Myr. Bottom: comparison between an extended radius of $\beta = 1$ and the formulation by \citet{salz-2016}, ran using the tracks for planets c (green) and d (blue). An efficiency of $\eta=15\%$ was used.\label{fig:evo}}
\end{figure}

We also evolved forwards two synthetic tracks for a planet with a core identical to planet b and envelope mass fractions equal to those of planets c and d at 10 Myr (as computed above). These tracks are evolved from the age of 10 Myr up to 700 Myr. The results show that, whichever envelope planet b had, it was likely lost within 5 to 10 Myr. This is expected, as b is the smallest and closest planet, thus more susceptible to photoevaporation. Planet b is therefore less useful in constraining the evaporation pasts of the system. Our evolutionary tracks are shown in Figure \ref{fig:evo} (top panel).

We also compare these results with tracks generated using an XUV radius that follows the relation by \citet{salz-2016} (Figure \ref{fig:evo}, bottom panel). In both cases, an efficiency of 15\% is used. We note a mass loss enhancement 2 to 5 times greater. This results in planets c \& d having a remarkably similar envelope mass fraction of about $\sim$0.9\% at 10 Myr.

\section{Discussion \& Conclusion}

Using \xmm{} measurements, we estimate the X-ray luminosity of K-type star K2-136 to be $1.18 \pm 0.1 \times 10^{28}$ erg s$^{-1}$ in the range 0.15 to 2.4 keV. The star, with an age of $700\pm100$ Myr, is about 50\% X-ray fainter than the luminosity predicted by \citet{johnstone-2020} in their stellar XUV tracks (see Figure \ref{fig:tracks}). The discrepancy is not unexpected, however, given the relatively large scatter of X-ray luminosities in relation to age \citep{jackson-2012, johnstone-2020}. 

We modelled the evaporation pasts of the three planets in the system using the stellar tracks by \citet{johnstone-2020}, the energy-limited mass loss formulation \citep{lecavelier-des-etangs-2007, erkaev-2007}, and the thermal evolution model by  \citet{lopez-fortney-2014}. We find that planets b and d are most likely rocky, given their current density and size, and planet c possesses a gaseous envelope that consist of 0.8\% of its total mass. Using the extended radius formulation by \citet{salz-2016}, we obtain similar initial envelope mass fractions for planets c \& d of 0.9\% at 10 Myr. \citet{rogers-owen-2021} finds the distribution of envelope mass fractions to be strongly peaked at 4\% for their sample of exoplanets, which might be attained by evolving our planets further back in time to 1 Myr with a alternate thermal evolution formulation.

Further mass loss might be attained by taking into account additional mass loss sources, such as stellar wind and Coronal Mass Ejections (CMEs). The degree to which these phenomena influence the atmospheres of mini-Neptunes is still under study. \citet{kislyakova-2014} found it to be several times lower than thermal mass loss rates. Fast CMEs, however, could have a greater influence on planets that are extremely close-in (<0.1 AU), or lack a significant magnetosphere \citep{lammer-2009}.
\blfootnote{The planetary evolution code used in this work is available on GitHub at \url{https://github.com/jorgefz/photoevolver}}

\appendix

\bibliography{main}

\end{document}